\documentclass[10pt,letterpaper]{article}
\usepackage{opex3}

\begin{document}

\title{Optical damage threshold of Au nanowires in strong femtosecond laser fields}

\author{A.M. Summers$^{1,4}$, A.S. Ramm$^1$, Govind Paneru$^2$, M.F. Kling$^{1,3}$, B.N. Flanders$^2$ and C.A. Trallero-Herrero$^{1,4}$}
\address{$^1$J.R.Macdonald Laboratory, Physics Department, Kansas State University, Manhattan, KS 66506, USA\\
$^2$Physics Department, Kansas State University, Manhattan, KS 66506, USA\\
$^3$Physik Department, Ludwig-Maximilians-Universit\"{a}t, 85748 Garching, Germany\\$^4$ Email: asummers@phys.ksu.edu, trallero@phys.ksu.edu}



\begin{abstract}
Ultrashort, intense light pulses permit the study of nanomaterials in the optical non-linear regime, potentially leading to optoelectronics that operate in the petahertz domain. These non-linear regimes are often present just below the damage threshold thus requiring the careful tuning of laser parameters to avoid the melting and disintegration of the materials. Detailed studies of the damage threshold of nanoscale materials are therefore needed. We present results on the damage threshold of Au nanowires when illuminated by intense femtosecond pulses. These nanowires were synthesized with the directed electrochemical nanowire assembly (DENA) process in two configurations: (1) free-standing Au nanowires on W electrodes and (2) Au nanowires attached to fused silica slides. In both cases the wires have a single-crystalline structure. For laser pulses with durations of 108 fs and 32 fs at 790 nm at a repetition rate of 2 kHz, we find that the free-standing nanowires melt at intensities close to 3 TW/cm$^2$ and 7.5 TW/cm$^2$, respectively. The Au nanowires attached to silica slides melt at slightly higher intensities, just above 10 TW/cm$^2$ for 32 fs pulses. Our results can be explained with an electron-phonon interaction model that describes the absorbed laser energy and subsequent heat conduction across the wire.
\end{abstract}


\section{Introduction}
\label{Introduction}
Nanoelectronics, nanophotonics and nanoplasmonics have been fields of growing interest in the last several years\cite{Ozbay,nanowire_quantum_dots,Quantum_plasmonics, Savage,Stockman}. These areas bring more conventional and explored fields in optics and electronics into the nanoscale regime, where many new phenomena occur and macroscopic descriptions break down or have to be strongly modified. Research into ultrafast nanoscale physics has great potential to both increase the fundamental understanding between light and matter and as lead to groundbreaking new technologies. As one such example, light-controlled nanoscale circuitry brings together all three of these sub-fields and opens the door to the next generation of optoelectronics-based devices that may operate at frequencies reaching into the petahertz domain \cite{Light_on_wire,Schiffrin_CEP_SiO2,Schultze_CEP_SiO2}.

Ultrafast nanoscale circuitry relies on the ability to utilize the non-linear interaction of light with nanoscale materials to tailor (collective) electron motion on a sub-optical-cycle timescale. These non-linear regimes are reached for strong laser pulses with intensities typically just below the damage threshold of the materials. Such interactions can give rise to (sub-optical-cycle) electron emission and acceleration from isolated nanotips \cite{krueger2011,herink2012} and nanospheres \cite{zherebtsov2011} and from nanostructured surfaces \cite{dombi2013,nagel2013}, the laser-field-driven semi-metallization of dielectrics \cite{durach2011,apalkov2012} and metals \cite{apalkov2012_metals}, and can induce currents across nanoscale junctions \cite{Schiffrin_CEP_SiO2}. In all these cases, the highest laser intensity that can be applied to the material depends on parameters such as material composition and quality and the pulse duration. The exact reasons for the damage of nanoscale materials are therefore often not well understood. Here, we study the melting of conducting, single-crystalline nanowires, which are one of the building blocks of nanoscale circuitry, under the illumination of femtosecond light fields with different intensities and pulse durations. Two types of nanowire arrangements are studied: (1) free-standing Au nanowires that are grown on W needle electrodes and (2) Au nanowires attached to silica slides.

Our investigations differ from previous damage threshold studies in a major way. Most studies of optically-induced damage and femtosecond heating in gold nanosystems have been performed for spherical nanoparticles embedded in a matrix or liquid \cite{Inouye,Grua2003,Baffou2011, Del_Fatti}, whereas we deal with free-standing single-crystalline nanowires. Other recent studies \cite{Liu2013} have analyzed the post damage structure of Ag nanowires after their laser-induced melting, but did not report the actual damage threshold. Since the composition of the nanosystem may play a major role in limiting the damage threshold, we employ single-crystalline nanowires that were grown with the directed electrochemical nanowire assembly (DENA) process \cite{DENA}. The low repetition rate of the laser used in our studies, with a spacing of 500 $\mu$s between individual pulses prevents the heating process from accumulating over many laser shots. Model simulations carried out for the interaction of a single laser pulse with a nanowire are in good agreement with the experimental observations.

\section{Experimental}
\label{Experimental}

\subsection{Experimental Setup}
\begin{figure}[htb]
\centering\includegraphics[width=9cm,clip=true,trim=0mm 0mm 0mm 1mm]{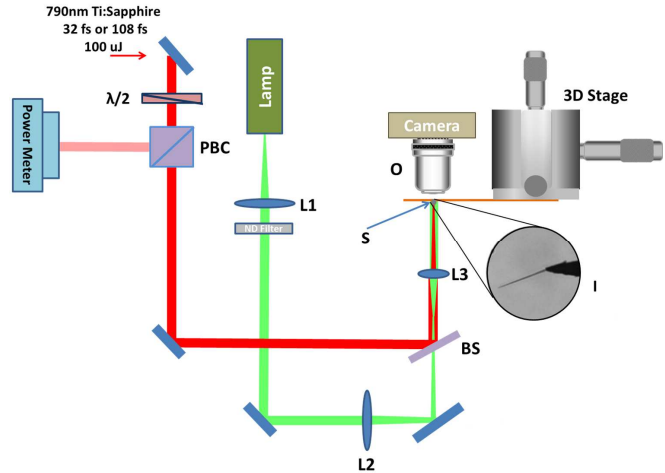}
\caption{Experimental setup for measuring the optical damage threshold of single-crystalline gold nanowires. $\lambda{}$/2, half-wave plate; PBC, polarizing beam cube; L1, 100 mm biconvex lens; L2, 100 mm biconvex lens; BS 50/50 beam splitter; L3, 60 mm plano-convex lens; S, sample mounting point; O, 20$\times$  long working distance objective; I, image of typical sample.   \label{fig:Nano_wire_set_up}}
\end{figure}

Our experiments made use of the Kansas Light Source (KLS), a chirped pulse amplification (CPA) \cite{Strickland} Ti:Sapphire laser system.  This laser provided pulses of 32 $\pm$ 3 fs in duration (FWHM of the intensity), with 1.8 mJ of energy and a central wavelength of 790 nm at a repetition rate of 2 kHz.  The pulses were attenuated to 100 $\pm{}$ 10 $\mu{}$J before entering the experimental set-up shown in Figure~\ref{fig:Nano_wire_set_up}. A half-wave plate and a polarizing beam cube were used to attenuate the peak power of the pulses while keeping the focusing conditions constant. The beam was focused onto the sample through ambient air, using a plano-convex 60 mm lens (L3), to a diameter of 9 $\mu$m (1/e of the intensity spatial profile). A manual three-dimensional translation stage was used to position the sample in the focus. The nanowires were sufficiently long to ensure that the laser focus illuminated only the wire and not the adjacent electrode.  A halogen lamp was used as a microscope backlight for visualization and alignment. We used lenses L1 and L2 as a condenser for the halogen lamp light as shown in Figure~\ref{fig:Nano_wire_set_up}. The halogen light was collimated by placing L2 at $\ f_{L2} + f_{L3}$ (the respective focal lengths) in front of L3, allowing for illumination of the entire sample.  Both the halogen and the laser radiation were imaged using a 20$\times$ microscope objective  (O) attached to a camera. In order to find the optical damage threshold, the samples were initially exposed to laser intensities significantly below the damage regime. To ensure that no optical damage was sustained, the sample was exposed to approximately 20,000 laser pulses (10 seconds) before moving to the next intensity step. The energy of the pulses was increased in 2.5 nJ steps until the first signs of damage were observed. Using the KLS grating-pair-based compressor, we were able to compensate for the frequency dispersion of all elements in the optical path.  The presence of a transform limited (TL) pulse in the interaction region was confirmed by measuring the pulse duration after the sample.  The pulse duration measurements were performed with all the optical elements in place using a Frequency Resolved Optical Gating (FROG) device.  In addition to the TL measurements, we also performed experiments with 108 fs, positively chirped pulses. These pulses were created by adjusting the grating distance in the amplifier compressor.  The relation between the laser polarization and the nanowire orientation was random.

\subsection{Sample Preparation}
The Au nanowires were fabricated by a technique called directed electrochemical nanowire assembly (DENA) \cite{DENA,Flanders}. This technique, based on dendritic solidification \cite{Nash,Kessler}, permits relatively straight-forward fabrication of metallic nanowires.  The diameter of these wires is tunable across the 20 nm to $\sim$ 1 $\mu$m range. The crystal structure of these wires is invariant along their lengths; hence they are single-crystalline \cite{DENA,Talukdar}.  We have employed two different electrode types to prepare Au nanowire samples.  The first set consists of nanowires fabricated at the tip of movable tungsten electrodes \cite{Paneru}. The set-up consists of two electro-etched tungsten electrodes immersed in \textit{HAuCl$_{4}$} solution. After mounting the tungsten electrodes on two different 3D stages and positioning them $\sim 1 \mu$m above a microscope slide, a 20 $\mu$l aliquot of aqueous solution containing 20.0 mM \textit{HAuCl$_{4}$} (Sigma Aldrich) was deposited across a $\sim$ 30 $\mu$m inter-electrode gap.  A function generator (Hewlett Packard, 8116A) was used to apply a square wave voltage signal of $\pm$ 4.0 V, 20.0 MHz to the tungsten electrodes to induce growth of the Au nanowire from the biased electrode towards the grounded electrode.  The voltage signal was turned off once the Au nanowire reached the desired length.  The wire was removed from the growth solution by translation of the microscope stage and allowed to dry. The second set consisted of Au nanowires that were grown at the tips of electrodes that were fabricated on a fused silica slide by evaporative deposition.  Roughly, a $\sim$  5$ \mu$m layer of gold was deposited on the glass slide using a vacuum evaporator (Varian VE10).  After growing the Au nanowire from the tip a of gold electrode by the method described above, it was twice cleaned by depositing a $\sim$ 10 $\mu$l  aliquot of deionized water on the wire, then wicking away the excess solution with a Kimwipe. Thus, the Au nanowires grown in this way lay in direct contact with the silica substrate.

\begin{figure}[htb]
\centering\includegraphics[width=9cm] {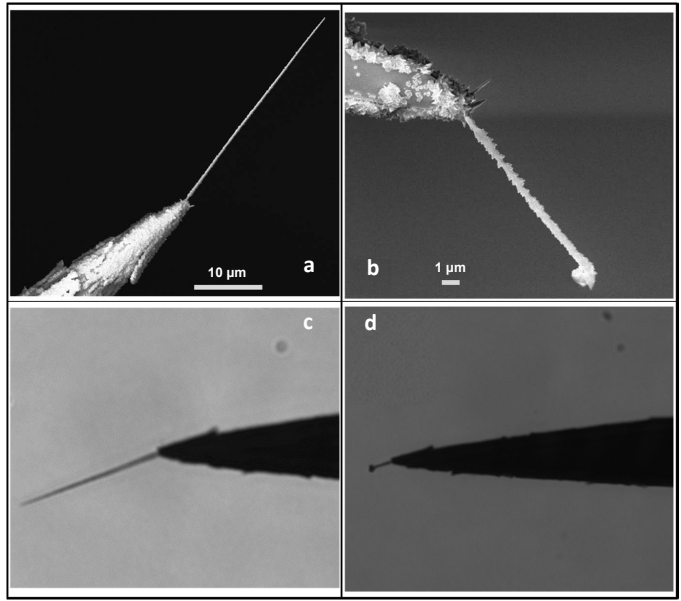}
\caption{Au nanowire grown free standing off of a tungsten electrode. (a) SEM image of a 25 $\mu$m long undamaged wire  (b) SEM image of a post exposure damaged wire (c) Same from (a) imaged in the experimental setup shown in Fig. \ref{fig:Nano_wire_set_up}. (d) Damage of nanowire from (a) and (c) as seen in experimental setup.\label{fig:Nano_wire_pics}}
\end{figure}

\subsection{Sample characterization and imaging}
Figure~\ref{fig:Nano_wire_pics} shows four images of typical nanowires used in this experiment.  Figure~\ref{fig:Nano_wire_pics} (a) shows a Scanning Electron Microscope (SEM) image taken before the sample was damaged.  The Au nanowire pictured is approximately 250 nm wide and 25 $\mu{}$m long, giving an aspect ratio of 100.  It was prepared using the DENA technique described above.  Figure~\ref{fig:Nano_wire_pics}  (c) shows the same nanowire, also prior to experimental damage, viewed using the experimental setup shown in Figure~\ref{fig:Nano_wire_set_up}.  The objective used was a 20$\times$ long-working-distance microscope objective.  A CMOS camera sensor with 2.2 $\mu{}$m sized pixels collected the image. Figure~\ref{fig:Nano_wire_pics} (d) shows the same nanowire after 5 seconds of exposure to a pulsed laser intensity of 8.8$\times 10^{12}$ W/cm$^2$.  Figure~\ref{fig:Nano_wire_pics}  (b) displays an SEM image of a typical nanowire post experimental damage under similar experimental conditions as the sample depicted in Figure~\ref{fig:Nano_wire_pics} (d). In total, we used 63 samples, of which 18 were free-standing on W electrodes, generated using the DENA process described above, and 45 were on fused silica, synthesized using the evaporated gold electrodes also described above. All samples were exposed to either 32 fs or 108 fs intense laser pulses.

\section{Experimental Results}

\begin{figure}[htb]
\centering\includegraphics[width=6.5cm,clip=true,trim=0mm 0mm 0mm 1mm] {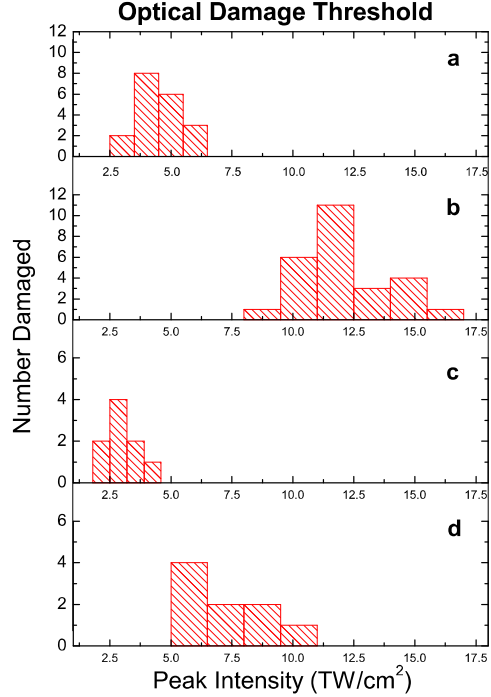}
\caption{Measured optical damage threshold distribution as a function of intensity for (a) 108 fs pulses and nanowires on fused silica slides, (b) 32 fs pulses and nanowires on fused silica slides, (c) 108 fs pulses and free standing nanowires on tungsten electrodes, and (d) 32 fs pulses and free standing nanowires on tungsten electrodes.   \label{fig:Damage_Data}}
\end{figure}

Our main experimental results are shown in Figure~\ref{fig:Damage_Data}. The figure shows the number of samples that were damaged at a given peak laser intensity in a histogram representation.  The damaging intensity was calculated from the energy measured when damage occurred by $I_{0} = \frac{4 E_{0}}{\pi w_{0}^{2} \tau} \sqrt{\frac{ln(2)}{\pi}}$ where $I_{0}$ is the peak intensity, $E_{0}$ is the pulse energy. $w_{0}$ is the focus radius, and $\tau$ is the pulse duration. The intensity bins used were 1 TW/cm$^{2}$, 1.5 TW/cm$^{2}$, 0.7 TW/cm$^{2}$ and 1.5 TW/cm$^{2}$ for Figure~\ref{fig:Damage_Data} (a),(b),(c) and (d), respectively. Figure~\ref{fig:Damage_Data} (a) and Figure~\ref{fig:Damage_Data} (b) show the distribution of damaged wires when these are on a fused silica substrate for pulse durations of 108 fs and 32 fs, respectively. Figure~\ref{fig:Damage_Data} (c) and (d) show the distribution of damaged wires when these are free-standing on a tungsten electrode for pulse durations of 108 fs and 32 fs, respectively. The influence of pulse duration becomes evident by comparison of Figure~\ref{fig:Damage_Data} (a) and (b) or Figure~\ref{fig:Damage_Data} (c) and (d). The influence of the growth process and heat conduction between the wires and the environment can be extracted from comparison of Figure~\ref{fig:Damage_Data} (a) and (c) for damage at 108 fs and Figure~\ref{fig:Damage_Data} (b) and (d) for damage at 32 fs. While the horizontal axis in Figure~\ref{fig:Damage_Data} is the peak intensity of the laser, it is also useful to compare when damage occurs as a function of pulse energy.  Rearranging the above equation for pulse energy and inserting all constants ($w_{0} = 4.5 \mu$m) gives $E_{0} = 3.7 \times 10^{-24} $(m$^{2}$s)$  I_{0}$ for 108 fs pules and $E_{0} = 1.1 \times 10^{-24} $(m$^{2}$s)$  I_{0}$ for 32 fs pulses.  This allows for the comparison of damage threshold in the energy picture.

Perhaps the most dramatic aspect to this measurement is the very high intensities in which the nanowires are found to survive. With average damage thresholds ranging from a few TW/cm$^{2}$ to above 10 TW/cm$^{2}$ for the respective cases, it is clear that Au nanowires survive well into the optical strong-field regime. In addition to this result, a clear distinction in optical damage threshold between the 32 fs and 108 fs cases can be seen for both wires grown on tungsten needles and on gold coated silica slides.  The damage threshold for both cases was higher for 32 fs pulses than for 108 fs pulses. This corresponds to a factor of $2.4 \pm 0.7$ for free standing nanowires and a factor of $2.7 \pm 0.6$ for the wires attached to fused silica substrates, where the uncertainty is due to the spread in the measured damage threshold. In contrast to this peak intensity picture, analysis of same data using the energy per pulse at which damage occurs we find only a small difference between all four cases. The average pulse energy at which the free nanowires damaged was 80 nJ for 32 fs pulses and 112 nJ for 108 fs pulses.  The wires attached to fused silica exhibited average damage pulse energies of 132 nJ and 169 nJ for the 32 fs pulses and 108 fs pulses, respectively.

We will first try to interpret these results from a simple thermodynamics point of view.  The first and simplest approximation is to consider the amount of energy needed to bring the nanowire to the melting temperature from room temperature.  We start with a textbook calculation of $Q$, the heat necessary to bring the gold wires to a melting point, $Q  = mc\Delta T$, with $m$ the mass of the nanowire, $c$ the specific heat of Au and  $\Delta T$ the change in temperature. To calculate the mass of the nanowire, we take into account the spot size of the laser focus $w_{0}=4.5 \mu$m and $d=150$ nm as the diameter of the nanowire (approximately the average diameter of our set of nanwires) . With these assumptions the amount of energy to heat the part of the nanowire in the laser focus up to it's melting point is $Q=1.7$ nJ

To compare this energy with the energy deposited into the wire, we take into account the percentage of each laser pulse that is incident on the wire by integrating over the overlap between the nanowire and the laser focus. We find that it takes a pulse of 62 nJ to bring the nanowire up to melting point ($\sim 2.7\%$ of the optical energy is incident on the nanowire).  Experimentally we find that pulses with a total energy of $\sim$ 61.5 nJ start damaging the wire. Of this total energy, $\sim$ 1.6 nJ is deposited on the wire. Thus, this very simplistic calculation agrees quite well with the experimentally observed amount of energy needed to melt the wires. Such agreement indicates that thermal exchange between the wires and the environment can be neglected and also that individual pulses, and not a cumulative heat effect from pulse to pulse, are responsible for the damage.

\section{Numerical Simulations}
A more sophisticated description of fs heating in coinage metal nanosystems breaks the absorption of optical energy and subsequent heating of the nanosystem into a 3-step process \cite{Grua2003,Baffou2011,Pulses1995}.  The first step, (1) {\it electronic absorption}, occurs when the pulse energy is absorbed by the free electron gas in the nanowire.  This occurs while the envelope of the field is present and results in a state where the electronic temperature has been dramatically increased while the temperature of the lattice has remained unchanged \cite{Baffou2011}.  The second step, (2) {\it electron-phonon thermalization}, is the cooling of this electron gas by interaction with the lattice phonons.  The characteristic time scale for this process ranges from a few picoseconds to a few hundreds of picoseconds depending on the system \cite{Baffou2011}.  The third step, (3) {\it external heat diffusion}, is the non-equilibrium thermal exchange between the nanosystem and the environment in which they are embedded.  In general, the third step is dependent on the geometry of the nanostructure and the environment but is often several orders of magnitude or more slower than the electron-phonon thermalization step. This last step is highly dependent on the type of environment the samples are embedded in.

All three of these steps are modeled by the following set of coupled differential equations \cite{Inouye,Easley,Lin2008,Anisimov}.
\begin{eqnarray}
\label{eq:general_propagation_1}
C_{l} \left(\frac{\partial T_{l}}{\partial t} \right) &= &C_{l}  \kappa_{l} \nabla^{2}(T_{l}) + G(T_{e} - T_{l}) -S_{l a}(\textbf{r}, t)  \\ 
C_{e} (T_{e}) \left (\frac{\partial T_{e}}{\partial t} \right) &=& C_{e} \kappa_{e} \nabla^{2}(T_{e}) - G(T_{e} - T_{l}) + S(\textbf{r}, t),
\label{eq:general_propagation_2}
\end{eqnarray}
where $T_{l} = T_{l}({\textbf r},t)$ is the lattice temperature profile, $T{e} = T_{e}({\textbf r},t)$ is the electron temperature profile, $C_{l}$ and $C_{e}$ are the respective lattice and electron heat capacities, $\kappa_{l}$ and $\kappa_{e}$ are the respective thermal diffusivities, $G$ is the electron-phonon coupling constant, $S_{l a}({\bf{r}},t)$ is the heat loss from the lattice to the surrounding air and $S({\bf{r}},t)$ is the absorbed power density from the laser pulse.

These equations, in their full form, are somewhat challenging to solve exactly.  Using physical insight to make several approximations, we can break these equations up into a decoupled set of more manageable equations.  First, our samples are always in air, and because the heat conductivity of gold is much larger than that of air or fused silica, a very small percentage of the heat will dissipate from the gold lattice to the surrounding air. Therefore, we will assume that there are no losses from the wires to the environment or $S_{l a} = 0$ .  The second approximation is that the thermal dissipation of the heated lattice will occur on a timescale that is several orders of magnitude slower than the electron heating or electron-phonon thermalization.  In this approximation all heating of the lattice has already occurred before any significant heat diffusion occurs.  This effectively lets us decouple the spatial term, $\kappa_{l} \nabla^{2}(T_{l}({\bf{r}},t))$, in Eq. \ref{eq:general_propagation_1} from the electron temperature by using the heat introduced from the electron-phonon interaction, $ G(T_{e} - T_{l})$, as the initial boundary condition, $T_{l}({\bf{r}},0)$, for the diffusion equation, Eq. \ref{eq:lattice_only}.  Also, because of the very large difference between the length and thickness of the nanowires, we need only solve the spatial equations in one direction.  Under these assumptions the heat dissipation of only the lattice in one dimension becomes
\begin{equation}
\frac{\partial T_{l}(x,t)}{\partial t} = \kappa_{l} \nabla ^{2}(T_{l}({x,t)}),
\label{eq:lattice_only}
\end{equation}
where $x$ is defined to run along the length of the nanowire, the other parameters remain the same as in Eq. \ref{eq:general_propagation_1} and $\kappa_{l} = 127 \times 10^{-6}$ m$^{2}$/s \cite{Benenson}. To find the spatiotemporal temperature evolution of the lattice $T_{l}(x,t)$ in our nanowires, we performed a one-dimensional finite difference numerical simulation of Eq. \ref{eq:lattice_only} for the geometry under question.

\begin{figure}[htb]
\centering\includegraphics [width=7.5cm] {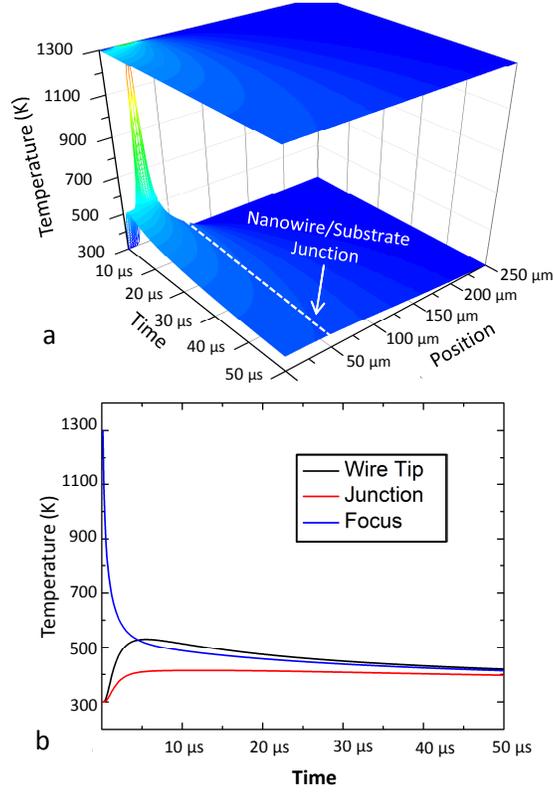}
\caption{Au nanowire cooling behavior from an initial Gaussian heat distribution with a peak temperature of 1300 K and a FWHM of $9 \mu$m (a) Temperature cooling profile as a function of both space and time in a 50 $\mu$m long Au nanowire attached to a conducting electrode. (b) Temperature line-outs of different points along the nanowire showing their temperature as a function of time after heating by a fs laser pulse.   \label{fig:Lattice_cooling}}
\end{figure}

The initial electronic excitation and subsequent thermal response occurs in a very localized region of the nanowire, defined by the size of the laser focus. To model this we assume an initial spatial gaussian temperature profile ($T_{l}(x,0)$) in the nanowire with $T_{max}$ = 1300 degrees K (just below the melting temperature of Au) and a full width at half maximum (FWHM) equal to the laser focus.  While this profile may not perfectly reflect the true initial temperature profile in a nanowire after interaction with a fs laser pulse, the important information in this simulation comes from analyzing the ensuing evolution of the temperature profile. The nanowires used in this experiment had strong thermal coupling to the bulk at the junction of the nanowire and electrode. To take this into account, we perform this simulation on a numerical grid that is approximately 10 times the size of the nanowire and place the initial temperature peak near the end of this sample grid.  Figure~\ref{fig:Lattice_cooling} (a) shows a surface plot depicting the transformation in the nanowire after being heated by a laser pulse.  The dashed line indicates where the nanowire would end and the electrode would begin (nanowire junction).  Figure~\ref{fig:Lattice_cooling} (b) shows line outs from this simulation at several key points of the nanowire (laser focus, nanowire tip and nanowire junction).  Analysis of this simulation shows that the temperature in the center of the focal point decays to approximately 20 $\%$ of the peak temperature within 5 $\mu$s.  50 $\mu$s after the pulse the entire nanowire has cooled to less than 100 degrees above room temperature, and when the next laser pulse comes, 500 $\mu$s later, the wire is only 20 degrees above room temperature (2 $\%$ of the maximum simulated temperature). This result shows just how slow the heat dissipation is and justifies separating this behavior from the general model given in Eq. \ref{eq:general_propagation_1}.  In addition this simulation shows that it is a reasonable assumption that there are negligible residual thermal effects in the nanowire for each subsequent pulse.

The heating of the electrons by the laser pulse and electron-phonon thermalization happens much faster than heat diffusion. At these timescales very little heat diffuses out from the heating area and thus $\kappa_{l} \nabla^{2}(T_{l}({\bf{r}},t)) \approx 0$. After removing this spatial contribution to the lattice temperature, $T_l,$  Eqs. \ref{eq:general_propagation_1} and \ref{eq:general_propagation_2} can be expressed as
\begin{eqnarray}
C_{l} \left(\frac{\partial T_{l}}{\partial t}\right) &= &  G(T_{e} - T_{l}) \\
C_{e} (T_{e}) \left(\frac{\partial T_{e}}{\partial t}\right) &=&  \kappa_{e} \nabla^{2}(T_{e}) -  G(T_{e} - T_{l}) + S({\bf{r}},t),
\end{eqnarray}
where the term $S(z,t)$ is a source term given by \cite{Easley},

\begin{equation}
 S(t,z) = (1-R) \alpha \hspace{1 mm}  exp( -\alpha z) I_{0}(t)
\end{equation}
with $R$ being the reflectivity of gold (0.974 at 800\,nm \cite{Palik}), $\alpha$ is the absorption coefficient (8.038 $\times 10^{5} cm^{-1}$ at 800nm \cite{Palik}) and $z$ is defined to be in the direction of the laser propagation.

In order to completely decouple the temporal derivatives from the spatial ones, we model the heating of the electron gas to occur in a fixed volume.  This "effective volume" is defined by the focal spot size, the nanowire radius, and an effective heating depth $z_{\rm{eff}}$ which removes the dependence on $z$ from the electron heat distribution $T_e(t)$ and the source term $S(t)$. Physically, the adoption of an effective depth implies that we will calculate the electron temperature distribution assuming that the wires heat uniformly up to a thickness $z_{\rm{eff}}$. In practice, $z_{\rm{eff}}$ is a fitting parameter to our source term and thus to our simulations.

This final approximation leaves us with our final set of equations to model the electron gas heating and electron-phonon thermalization steps, widely used in the literature to explain the heating of nanoparticles \cite{Inouye,Grua2003,Baffou2011,Pulses1995,Lin2008,Anisimov}.

\begin{eqnarray}
C_{l} \left(\frac{\partial T_{l}}{\partial t}\right) &= & G(T_{e} - T_{l}) \\
C_{e} (T_{e}) \left(\frac{\partial T_{e}}{\partial t}\right) &=& - G(T_{e} - T_{l}) + S(t),
\label{eq:temperature_effective}
\end{eqnarray}
and
\begin{equation}
 S(t) = (1-R)\alpha \hspace{1 mm}  exp(-\alpha z_{\rm{eff}}) I_{0}(t),
\label{eq:source_effective}
\end{equation}
where $C_{e} \approx \gamma T_{e}, \gamma = 63$ J\,m$^{-3}$K$^{-2}$ \cite{Grua2003}, $C_{l} = 2.49\times 10^{6}$ J $m^{-3} K^{-1} $\cite{CRC_Handbook}, and $G = 2.5\times 10^{16}$ W\,m$^{-3}$K$^{-1}$ \cite{Fann1992}.  Equations \ref{eq:temperature_effective} and \ref{eq:source_effective} represent a phenomenologically effective model that does not completely model the true electron heating dynamics from an intense, ultrafast source. Instead, this model assumes the formation of a hot electron gas with a mean value given by $T_e(t)$. In our simulations, $z_{\rm{eff}}$ is used as the single fitting parameter to match the simulated damage threshold (defined as the lowest intensity required to raise the nanowire to the melting temperature of gold) to the data shown in Figure~\ref{fig:Damage_Data}.

\begin{figure}[htb]
\centering\includegraphics [width=7.5cm]{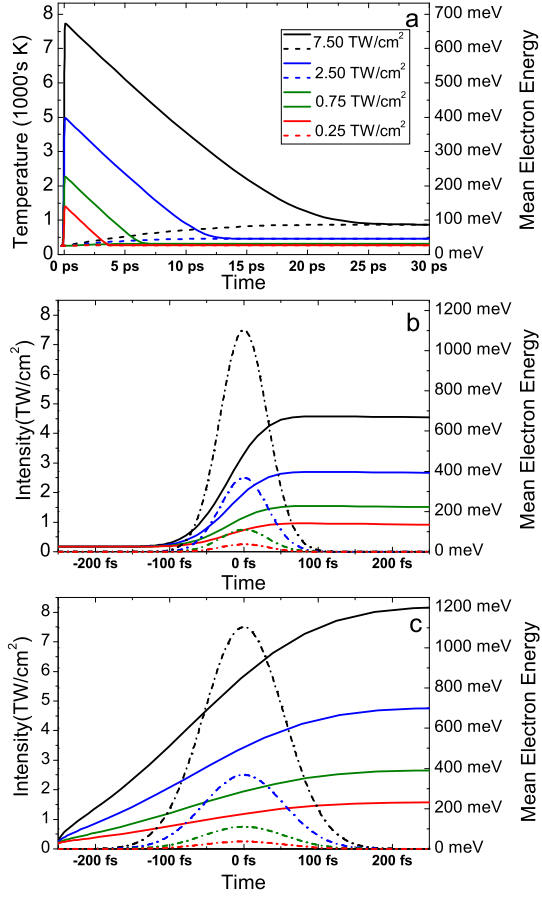}
\caption{(a) Temporal profile of both lattice temperature (dashed lines) and mean electron energy (solid lines) after interaction with 7.5\,TW/cm$^2$, 2.5\,TW/cm$^2$, 0.75\,TW/cm$^2$ and 0.25\,TW/cm$^2$ 32\,fs laser pulse. Fast timescale electron energies (solid lines) are shown for (b) 32 fs pulses and (c) 108 fs pulses, along with the respective laser intensity temporal profiles (dot-dash lines).     \label{fig:Electron_Lattice_Sim}}
\end{figure}

By solving these equations numerically using a variable step fourth-and fifth-order Runga-Kutta method \cite{MATLAB}, we found both the electronic and lattice temperatures as a function of time after being illuminated with a femtosecond laser pulse. The results of these simulations are shown in Figure \ref{fig:Electron_Lattice_Sim}. Our simulations show that a factor of 3.3 increase in intensity is required for 32 fs over 108 fs pulses to heat the wire to the melting point.  This is comparable to the factors of 2.4 and 2.7 found experimentally for the tungsten needle and evaporated gold substrate samples, respectively.  Figure \ref{fig:Electron_Lattice_Sim} (a) shows the lattice and electron temperature over a long timescale for four separate peak laser intensities, 7.5 TW/cm$^{2}$, 2.5 TW/cm$^{2}$, 0.75 TW/cm$^{2}$, and 0.25 TW/cm$^{2}$.  Looking at both the electron and lattice temperatures as a function of time shows two very distinct time scales, thus confirming the above two-step model.  The figure also shows that the cooling time is dependent on the peak laser intensity and that only a small fraction of the electron initial temperature is transferred to the lattice. Figure \ref{fig:Electron_Lattice_Sim} (b) and (c)  show the electron temperature (given in units of mean electron energy) for a 32 fs pulse in (b) and 108 fs in (c).  These plots show that the heating happens on a time scale defined by the laser pulse duration, since the peak electron temperature is reached very shortly after the peak of the laser pulse. This observation is in agreement with attosecond resolved electron dynamics in solids \cite{Thum_Atto_Surfaces,cavalieri_atto_CM} and in the electron tunneling in atoms, \cite{ionization_delay_review} where the electrons released through the tunneling process almost instantaneously follow the laser field.

Comparing the different electron temperatures for the 32 fs and 108 fs in Fig. \ref{fig:Electron_Lattice_Sim} (b) and (c) and the experimental results in Fig. \ref{fig:Damage_Data}, it is clear that longer pulses induce greater heating into the system resulting in a lower damage threshold. This result is in agreement with strong-field ionization of atoms and molecules where the ionization saturation peak intensity is dependent on the laser pulse duration \cite{Hankin2001}. Qualitatively, this comes from the fact that the longer a pulse, the more likely an electron is to absorb photons and be excited into the conduction band. Therefore, a straight forward manner to use high intensities for laser-matter interaction is to reduce the pulse duration. Few-cycle pulses with bandwidth ranging from the ultraviolet to the mid-infrared are a present reality \cite{Hassan2012, Huang2012, Schmidt2012}. For this reason, we also performed calculations, using Eq. \ref{eq:temperature_effective}, for $T_l$ and $T_e$ using a 5 fs pulse. For such short pulses we find that damage occurs at peak intensities exceeding 50 TW/cm$^2$! It is however, anticipated that the adiabatic, phenomenological model used to arrive to Eqs. \ref{eq:general_propagation_1} and Eq. \ref{eq:general_propagation_2} cannot accurately describe the dynamics for such pulses. At peak intensities higher than $10^{13}$ W/cm$^2$ the fields are so strong that the material properties become field-dependent \cite{apalkov2012_metals}. The accurate modeling of the interaction of laser pulses with such high intensity furthermore requires a microscopic description of the laser-driven dynamics \cite{varin2012}.

\section{Conclusions}
We have presented measurements and simulations on optically-induced damage in Au nanowires with intense, femtosecond laser pulses. Our experimental measurements indicate that the damage threshold in single-crystaline nanowires, attached to silica, can reach values close to 10 TW/cm$^2$ for 32 fs pulses and approximately 5 TW/cm$^2$ for 108 fs. The dependence of the damage threshold peak intensity on pulse duration is similar to that observed in strong field ionization of atoms and molecules, where ionization saturation intensity depends on pulse duration. This is further confirmed by the fact that the pulse energy at which the damage occurs is similar for both pulse durations. Using coupled electron-lattice heat dissipation, we find that the peak temperature of the electron gas is reached shortly after the peak of the pulse and that this delay depends on the pulse duration. By extrapolating our calculations to 5 fs pulses, we speculate that, peak intensities of ~ 50 TW/cm$^2$ may be attained before melting occurs. Our findings provide critical information for all future studies of nanowires using ultrafast lasers. The fact that these systems survive well into the strong field regime shows that ultrashort light pulses permit studies in the non-linear regime, with many new opportunities for applications towards ultrafast light-driven nanoelectronics.

\section{Acknowledgement}
The JRML personnel acknowledges support by the U.S. Department of Energy under DE-FG02-86ER13491. M.F.K. acknowledges support by the U.S. Department of Energy under DE-SC0008146 and partial support by the DFG via the Cluster of Excellence: Munich Center for Advanced Photonics (MAP).
\\
\\
\noindent A.M.S. and C.T-H thank Kevin Carnes for the revision of the manuscript.

\end{document}